\documentclass[aps, prper, reprint,groupedaddress]{revtex4-2}  
\usepackage{graphicx, tabularx, longtable, nicefrac, amsmath,enumerate, supertabular, eurosym}
\usepackage{microtype}
\usepackage{booktabs}
\usepackage[utf8]{inputenc} 
\usepackage{csquotes}
\usepackage{booktabs}
\usepackage[normalem]{ulem}

\usepackage{hyperref,xcolor}

\begin{document}


\title{Educational data augmentation in physics education research using ChatGPT}



\author{Fabian Kieser$^1$}
\author{Peter Wulff$^1$}
\author{Jochen Kuhn$^2$}
\author{Stefan K{\"u}chemann$^2$}
\email{mailto: s.kuechemann@lmu.de}

\affiliation{$^1$ Physics and Physics Education Research, Heidelberg University of Education, Im Neuenheimer Feld 561, 69120 Heidelberg, Germany}
\affiliation{$^2$ Chair of Physics Education, Faculty of Physics, Ludwig-Maximilians-Universit{\"a}t M{\"u}nchen (LMU Munich), Geschwister-Scholl-Platz 1, 80539 Munich, Germany}


\date{\today}
\begin{abstract}
Generative AI technologies such as large language models show novel potentials to enhance educational research. For example, generative large language models were shown to be capable to solve quantitative reasoning tasks in physics and concept tests such as the Force Concept Inventory (FCI). Given the importance of such concept inventories for physics education research, and the challenges in developing them such as field testing with representative populations, this study seeks to examine to what extent a generative large language model could be utilized to generate a synthetic data set for the FCI that exhibits content-related variability in responses. We use the recently introduced ChatGPT based on the GPT 4 generative large language model and investigate to what extent ChatGPT could solve the FCI accurately (RQ1) and could be prompted to solve the FCI as-if it were a student belonging to a different cohort (RQ2). Furthermore, we study, to what extent ChatGPT could be prompted to solve the FCI as-if it were a student having a different force- and mechanics-related preconception (RQ3). In alignment with other research, we found that ChatGPT could accurately solve the FCI. We furthermore found that prompting ChatGPT to respond to the inventory as-if it belonged to a different cohort yielded no variance in responses, however, responding as-if it had a certain preconception introduced much variance in responses that approximate real human responses on the FCI in some regards.
\end{abstract}


\maketitle

\section{Introduction}
Conducting empirical educational research is at the center of learning and instruction sciences, for instance, for evaluating the effectiveness of instructional materials, proving theoretical frameworks, or designing test instruments. The design of effective learning materials or test instruments often requires a large amount of human effort to perform the studies and several iterations within the empirical validation process for the optimization of the materials. Additionally, conducting empirical studies requires financial investments, such as providing incentives for participation.

Furthermore, in the past decade, learning systems based on artificial intelligence (AI), such as AI-based intelligent tutoring systems or automated feedback systems, have received growing attention, and several efforts have targeted the implementation of AI tools in learning environments. For a good performance, such AI tools require training and fine-tuning with a large amount of empirical data. Although there are a growing number of pre-trained AI-based models that provide good results for cases similar to the training data, it is often necessary to fine-tune them for desired performance in specific applications.

Recent advances of large language models (LLM) could have several opportunities for education. Kasneci and colleagues (2023) suggest the number of opportunities for learning and teaching, such as a support in writing tasks, development of critical thinking, individual learning opportunities, semi-automated grading of students' works and lesson planning \cite{kasneci2023chatgpt}. For educational research, LLM may also help with summarizing and categorization tasks. For instance, van de Shoot and colleagues (2021) developed a tool for automated title and abstract screening for systematic literature reviews \cite{van2021open}, which was shown to save a significant amount of time. Particularly, the release of ChatGPT (based on GPT 3.5) has stimulated several research efforts to explore the opportunities for education. For instance, ChatGPT can be used summarizing texts, zero-shot text classification tasks and programming tasks \cite{kuzman2023chatgpt,surameery2023use}. 

It was also found that ChatGPT is able to solve a variety of problems with zero-shot learning, it performs reasonably well on university exams in medicine and can support in solving higher-order problems in pathology \cite{Sinha.2023}. These results offer the opportunity to test developed assessments before administering it to students and subsequently optimizing them based on the answers of ChatGPT. This is an important benefit, as it can save time and effort in the assessment development process and yield insights into psychometric properties of the instruments. Additionally, in science education research, concept tests, such as concept inventories, play an important role to diagnose students' conceptual understanding and inform teachers about prevailing students' difficulties. 

In physics education research, the most used concept inventory (CI) is the Force Concept Inventory (FCI) developed by Hestenes and colleagues \cite{hestenes1992force}. As a common concept test, it consists of several multiple-choice questions that target specific preconceptions related to  Newtonian mechanics, in particular the force concept. Therefore, the answer alternatives are specifically designed to reflect these preconceptions. To use AI-based methods such as LLM to test features of concept tests, e.g., if they capture students' difficulties, it is necessary that LLM are aware of these difficulties and know how students would answer as-if they experience one or more of these difficulties. Furthermore, if a LLM would be able to answer concept questions from the perspective of a student who experiences difficulties, the created synthetic data could be used to train and fine-tune smaller AI tools for education that support learners or better inform teachers. Therefore, it would be a great benefit if LLM are able to create a synthetic data set that closely resembles a realistic student data set, as it would save a significant amount of time and effort in the development of assessments and concept tests.

For this purpose, we explore in this work the opportunity of creating synthetic empirical data with LLM. First, we provide an overview of relevant previous works and then specify research questions that target certain aspects of the suitability to generate a synthetic data set in physics education research.

\section{Theoretical Background}

\subsection{Concept inventories in science subjects}\label{concept_inventories}
Physics in particular and science subjects more generally are characterized by a structured knowledge base that is centered around key concepts \cite{Koponen.2010,Wulff.2023}. Hence, the conceptual understanding of key concepts in scientific disciplines such as evolution or force is important, and key to scientific literacy. To measure high-level conceptual understanding and evaluate pedagogical practices, science education researchers devised so-called concept inventories \cite{Liu.2010,Porter.2014}. Concept inventories have been developed and widely employed to assess the understanding of key concepts such as biological evolution \cite{Nehm.2012}, climate change \cite{Schubatzky.2023}, or Newtonian force \cite{Hake.1998}. These concept inventories typically involve multiple-choice questions, where designing distractors that capture common preconceptions is of crucial importance. Concept inventories have helped science education researchers to empirically show to what extent a basic level of conceptual understanding is reached in school and university education. Oftentimes, it is found that a full understanding of key concepts such as in Newtonian mechanics the force concept is typically reached only by few students even after (traditional) instruction. Moreover, concept inventories helped researchers to glean evidence supporting the effectiveness of interactive engagement instructional formats, which yielded higher learning gains \cite{Hake.1998}. The concept inventories thus helped to substantially improve instruction, e.g., of mechanics in introductory physics, across universities \cite{Eaton.2021}.

\subsection{Development of CIs}\label{development_inventories}
Designing concept inventories was described by Adams and Wieman as a process consisting of consecutive phases \cite{Adams.2011}, namely 1) delineation of purpose and scope of the construct to be measured, 2) development and evaluation of test specifications, 3) field testing, evaluation, and selection of items, and 4) assembly and evaluation of the test. Phase 1 includes information about expert thinking on the construct and requirements for the concept inventory to be implemented, such as no required training and ease of use. In Phase 2, decisions on psychometric properties are considered, as well as the target population for which the concept inventory is designed. Phases 3 and 4 comprise the bulk of the work needed to design the concept inventory. Six iterative steps are distinguished that include a) establishing topics of interest, b) interviews and observations to identify student thinking, c) create open-ended survey to probe student thinking, d) create forced-answer test, e) carry out validation interviews, and f) administer to classes and run statistical tests. Establishing the topics of interests typically requires a resource-intense Delphi Process \cite{Porter.2014}. Often, 15 to 20 experts need to be recruited to discuss and refine topics or constructs in concept inventories. \cite{Porter.2014,Adams.2011}. Developing a concept inventory is thus an extensive and time-consuming task that involves substantial resources, where AI technologies might be useful in almost any phase and step. Once qualitative, open-ended analyses are completed, suitable forced-choice (e.g., multiple-choice) items have to be field-tested where students' responses are sampled in order to determine the underlying structure of the concept inventory in terms of psychometric properties \cite{Eaton.2021,Adams.2011}. In the field tests class sizes of a ``few hundred or more [students] are desirable'' \cite[p. 1303]{Adams.2011}. In line with generalizability theory, administering parallel tests of tests to two equivalent populations are desired \cite{Adams.2011}. However, these requirements are difficult to meet, and place restrictions on small institutions that might not be able to develop concept inventories in the first place. It might also be difficult to sample representative student populations, due to various reasons such as restricted access, ecological considerations, or conceptual problems such as defining the target population. Concept inventories would also have to be field tested to different age groups to determine potential differences \cite{Eaton.2021}. 

Given this resource intensive process of field testing concept inventories it would be desirable for the PER community to enhance existing strategies to field test CI. A recently evolved strategy might be the use of generative AI technologies such as LLM that were found to excel at various language generation tasks that even involved knowledge application. LLMs have been found to be capable of generating synthetic data and solving problems without being explicitly instructed to do so; therefore, they are valuable candidates for outsourcing some of the field testing in concept inventory development—and test instrument development more generally. 

\subsection{Large Language Models (LLM)}\label{LLMs}
AI technologies have enabled tremendous progress in fields such as computer vision and natural language processing (NLP). A first breakthrough moment for machine learning (inductive problem solving by computers) was probably when deep neural networks could be utilized to accurately classify pictures. Even more so, since its inception \cite{Bengio.2003}, and especially since 2017 so-called LLM have had a profound impact on solving all sorts of language-related problems as diverse as question answering, translation, or summarizing text \cite{Vaswani.2017,Devlin.2018,Manning.2022}. There is a vast variety of LLM architectures that accomplish specific tasks. For example, sequence-to-sequence models were used to translate an input to an output, or encoder architectures were used to classify an input into predefined categories. Most often, some form of deep neural networks underlie the architecture of LLM. A general feature for these LLM was that they are sensitive to number of parameters of the artificial neural network, train set size, or amount of compute time, where increases in all of these features generally improve the performance of the LLM \cite{Kaplan.2020}. 

Many LLM are trained in two steps. The first step consists semi-supervised learning in which a masked language model objective is pursued. In masked language modeling context words are masked out and have to be predicted by the algorithm, given a related sequence of words \cite{Wolfram.2023}.\footnote{More precisely, tokens are used instead of words, which are more general sequences of letters that can be added to form words. However, they have the advantage that also unseen words in the training data can be processed.} Early LLM were trained based on large text corpora such as the Common Crawl (i.e., dump of the Internet, 60\%\footnote{Percentages are taken from \url{https://en.wikipedia.org/wiki/GPT-3\#GPT-3.5} (last access 26 May 2023) for training GPT 3.}), WebText2 (22\%), or Wikipedia (3\%) with the training objective to predict next words in a sequence (a sort of Cloze test, \cite{Devlin.2018}), and classify if a subsequent sentence actually follows a given sentence \cite{Devlin.2018}. To the surprise of many experts, such a simple training objective enabled large enough models to gain capabilities to perform entirely novel tasks that were not seen in the training process \cite{Brown.2020}. Some resemblance of masked language modeling to human language processing and cognition, namely predictive coding where future words are sought to be predicted given the current words, was recognized \cite{Rosenbaum.2022}. The second step consists of a supervised learning task in which the pre-trained model (also called foundational model, \cite{Bommasani.2022}) is trained or fine-tuned on a specific task on a labeled data set. In this part, the LLM predicts classes based on textual information. 

Among the first LLM that utilized a transformer architecture and solved a wide range of unseen tasks was the Generative Pretrained Transformer 3 (GPT 3) by OpenAI. This model was tested on a wide range of tasks such as common sense reasoning. It was found that it achieved substantial accuracy \cite{Brown.2020}. A novel paradigm emerged, namely prompting LLM, i.e., providing it context with the input that it uses to generate a response. One prompting strategy was to provide the LLM examples of similar task solutions that could improve accuracy \cite{Brown.2020}. Some researchers contended that LLM with their tremendous size (GPT 3 has 175B parameters) overfit the training data and are essentially “stochastic parrots” \cite{Bender.2021} that regurgitate the training data. However, analyses of data contamination (i.e., evaluation examples being present in the training data) revealed that the LLM could solve tasks that were not contained in the training data set \cite{Brown.2020}. 

While word analogy problems and other fairly simple problem have been solved by LLM rather early, quantitative reasoning as in disciplines such as biology, chemistry, or physics was a more difficult task to solve. AI researchers introduced more advanced training tasks that enabled LLM to also solve quantitative reasoning problems as well to some extent \cite{Lewkowycz.2022}. In particular, chain-of-thought prompting was found to be particularly effective \cite{kojima2022large}. For once, researchers used science-related content to specifically further fine-tune (i.e., adapt the weights) the LLM in a learning paradigm called transfer learning. It was later found that generative LLM such as GPT required task-specific fine-tuning. However, AI researchers showed that scaling up LLM also enabled them to solve tasks they have not been specifically fine-tuned for, so-called few-shot or even zero-shot learning \cite{Brown.2020}. These LLM achieved translation, question-answering, and Cloze tasks with only seeing few or zero examples, and even accomplished reasoning-related capabilities \cite{Brown.2020,Huang.2023}. With increases in model size (GPT 3 had 175B parameters) LLM became increasingly capable to perform unseen tasks, such as two- or three-digit arithmetic. Interestingly, some capabilities only emerge with sufficient model size (number of parameters), compute steps during training (e.g., training epochs), and training data size \cite{Bowman.2023}. Nowadays, LLM are as flexible as to pass exams in medicine, law, or business without specific fine-tuning \cite{Katz.2023}. Also for concept inventories in physics, GPT- based LLM such as ChatGPT were found to achieve almost 100\% accuracy \cite{Kortemeyer.2023,west2023ai}. Given these findings, it is a next step to find out whether and how to utilize LLM-based tools such as ChatGPT for pilot-testing purposes for CIs.

\subsection{Data augmentation techniques}
The widespread use artificial neural networks is mainly attributed to their exceptional performance in certain classification and pattern recognition tasks. To achieve such high performance, the training of neural networks requires a large amount of data. However, in some cases, it is difficult to collect a large amount of data, such as empirical data in education or time series data \cite{iwana2021empirical}. This is critical, because a smaller number of training data usually leads to a reduced generalization ability and a lower overall performance of the model. One solution to increase the generalization ability of trained models by reducing the chance of overfitting is the generation of synthetic data, the so-called data augmentation \cite{van2001art}. 

In the field of image recognition, data augmentation is a common practice in which, for example, the original image is cropped, mirrored, or its colors are changed to enhance the number of images in the training set \cite{shorten2019survey}. In time series data, a recent literature review provides an overview of magnitude, time domain, and frequency domain transformation techniques \cite{iwana2021empirical}. In an additional empirical evaluation of various data augmentation methods for time series data, the authors report both enhanced and decreased generalization accuracy. For example, a permutation method in which the observations and time of the observations is changed, leads to detrimental effects, as it breaks the time dependence of the series.

In our case, the educational data to be enhanced are the solutions on a standardized concept inventory, the FCI. The analysis of data augmentation of educational data is important for test instrument development and validation as well as for adaptive learning systems that provide a personalized response to a learner's input or behavior. In concept tests, the items are related to the concepts, but there is no inherent sequence dependence. Often the items are even administered in a random order to avoid potential effects of the item sequence. Therefore, typical data augmentation techniques in the range of time domain transformations are not suitable for educational data. 
In addition, students' responses to concept test items are not three dimensional as images (x-coordinate, y-coordinate, color). Therefore, common techniques of image augmentation are not a promising augmentation method for educational data. 

One technique that was used across different fields of data augmentation is jittering, in which a small offset is given to the original data. However, educational data such as students' responses to concept test items are discrete and not continuous, and small variations in the responses or the item, in which a certain response was given, would lead to randomly incorrect responses. Consequently, this technique would have a similar effect as permutation techniques and it is not likely to perform well. 

Therefore, it is important not to translate these existing methods for educational data augmentation but to be sensitive to the inherent structure of the data and study options that consider the structure of concepts in the test items. In contrast to the data types mentioned above, educational data often exhibits a hierarchical structure, e.g., a student is in a specific class with a certain teacher and the class is part of a specific school. To account for this, it is important that a suitable augmentation technique for educational data would consider different levels of prior knowledge. As mentioned earlier, large language models have made a tremendous advance in the past years and they have shown to be able to respond reasonably well on physics concept test items. Therefore, they are a promising method to augment educational data, but it is an open question to what extent and how systematically a large language model can reflect different levels of prior knowledge and students' difficulties.

To assess the suitability for generating a valuable synthetic dataset in the field of education and to account for varying levels of prior knowledge, this study explores the following research questions:
\begin{enumerate}
\item[RQ1:] What is the conceptual understanding of the force concept of ChatGPT?
\item[RQ2:] To what extent is ChatGPT able to simulate conceptual understanding of different cohorts?
\item[RQ3:] Is ChatGPT able to simulate specific preconceptions of students?
\end{enumerate}

\section{Methods}
\label{sec:methods}
\subsection{Participants}
We compared the responses to the FCI given by ChatGPT with real students in the first semester at the RPTU Kaiserslautern-Landau, Germany. The engineering students attended a lecture of experimental physics covering the topic of mechanics, among others. The FCI was administered at the beginning of the semester, so at the moment of answering the test items, the students had only covered these concepts in high school but not yet during the lecture. In total, there were 121 engineering students (12 female, 105 male, and 4 made no statement) with an average age of 20.6 years.  

\subsection{The Force Concept Inventory}
To examine to what extent ChatGPT can be used fruitfully in educational contexts for tasks such as solving standardized test instruments and generate data, we evaluated performance in the domain of mechanics as it is coherently structured with regards to knowledge. Physics education researchers devised standardized test instruments that are widely employed across universities throughout the world. Among the most well-known instruments are concept inventories that probe learners' deep understanding of a particular concept. Physics educators established that learners come with intuitive conceptions of certain physics-related concepts, also referred to as common-sense knowledge \cite{Hammer.2000}. These might not coincide with more normative ideas about physics, such as Newtonian physics, and are typically context-dependent \cite{Hammer.2000}, Most learners developed over her/his upbringing an intuition of forces as acting agents \cite{hestenes1992force,diSessa.1993}, both by experiencing and interacting with the world and hearing language such as \enquote{gravity pulls}. These intuitive conceptions are largely incompatible with the rich physics conceptions of forces, such as in Newtonian mechanics where Newton's laws lay out a consistent conceptualization of forces. The first law states that an object at constant velocity is remaining in this state of constant velocity unless a force is acting upon it. The second law states that the acceleration of the object is dependent upon the force and mass of the object. Finally, the third law states that every force has an opposite equal force (actio equals reactio). 

To probe students' understanding of Newtonian mechanics, the FCI was developed \cite{hestenes1992force}, originally in 1992, and revised in 1995, containing 30 multiple-choice items. The FCI, as is typical for physics problems, combines image and text data. However, as ChatGPT based on GPT 4 only accepts text data as input, we transcribed the pictures in the FCI, similar to the procedure employed by West \cite{west2023ai}. For example, a typical FCI item would include a picture of the following scenario:

\begin{displayquote}
\enquote{A heavy ball is attached to a thread and moved horizontally around in a circle. At one point, the thread suddenly breaks. The whole process is viewed from above. Which trajectory best describes the path of the ball after the thread breaks? 1. a trajectory curve that nearly follows the path of the circle 2. a straight-line trajectory curve that points tangentially along the circle at the break point 3. a straight-line trajectory curve that points straight away from the circular path at an angle of 45° from the tangent at the break point 4. a trajectory curve that initially goes outward from the center, but then takes a trajectory curve in the direction of the original motion 5. a trajectory curve that points straight outward from the center.}\footnote{Chat can be accessed here: \url{https://chat.openai.com/share/8ae109d2-8b71-4033-8c62-749c33c2582c}, last access: June 2023; translated with: \url{https://www.deepl.com/de/translator#de/en/})}
\end{displayquote}

West \cite{west2023ai} verified that besides the transcriptions ChatGPT is still capable to solve the FCI. The transcription of all items that contained a figure can be found in the Appendix. 

\subsection{RQ1: Probing conceptual understanding of ChatGPT in physics}

Probing accuracy of ChatGPT on physics contents and the FCI was performed by multiple researchers \cite{west2023advances,Kortemeyer.2023,west2023ai}. Some studies examined qualitative physics-related dialogues with ChatGPT, and short-form physics essays \cite{Gregorcic.2023,Yeadon.2023}. Gregorcic and Pendrill \cite{Gregorcic.2023} found for a simple prompt (\enquote{A teddy bear is thrown into the air. What is its acceleration in the highest point?}) that ChatGPT's responses (based on GPT version 3.5) were full of contradictions and concluded that ChatGPT is not good enough as a cheating tool in physics. Yeadon et al. \cite{Yeadon.2023} used ChatGPT (based on version 3.5) to generate responses (300 word essays) to questions and found that it performed on par with the students. Plagiarism detection software was largely incapable to distinguish it from human writing. Kortemeyer \cite{Kortemeyer.2023} then used ChatGPT (version 3.5) on the FCI and found that it scored correctly 18 out of 30 questions (60\%). This would equal approximately the performance of a beginning learner. To assure that ChatGPT was not simply trained on the FCI (or at least could generalize beyond mere textual matching), Kortemeyer changed surface-level information in an item and found that ChatGPT was not distracted by these changes, and still performed correctly. It was also found that ChatGPT made several errors that related to the impetus (I) preconception. Moreover, other errors related to confusion of individual forces versus net force on an object, unstable concepts, and logical errors \cite{Kortemeyer.2023}. West \cite{west2023ai} could replicate these findings for ChatGPT based on GPT version 3.5, and found that a substantial improvement to expert level performance (22 correct out of 23) could be observed when using ChatGPT based on GPT version 4. In RQ1, we will generate a larger sample of response sets for ChatGPT (based on GPT version 4) and calculate the accuracy and standard deviation of the responses against the correct answers. This will enable us to estimate the degree to which ChatGPT is certain of its answers. Similar to West \cite{west2023ai}, we will then compare ChatGPT's distribution of the number of correct answers to the distribution of a student cohort.

\subsection{RQ2: Simulating conceptual understanding of different cohorts}

An important capability of LLM for education is to simulate (synthesize) data, given specific prompting strategies. In general, pre-trained LLM generate output based on the context provided in the input. The provided input context determines the output, given that a different sequence of tokens is to be predicted. It is then possible to adjust the response behaviour of ChatGPT for the FCI questions. For example, one could request that ChatGPT always answers randomly for each question. A more informative prompt would be to ask ChatGPT to answer from the perspective of a different student group, which could be related to background variables, such as experience. In RQ2, we chose the experience-related variable, and prompted ChatGPT to generate responses from the perspective of a) school students ($N$=8), and b) engineering students ($N$=18). We chose these groups to potentially generate variance and see to what extent differences emerge in ChatGPT's responses. We then generate for each cohort a number of response sets in order to analyze the distribution of correct answers, comparing the cohorts with the real students and the plain ChatGPT response sets. As a robust (non-parametric) test statistic, suited for multiple groups, we utilized Kruskal-Wallis rank test \cite{Kruskal.1952}. This test does not assume normality of residuals, though it assumes identically shaped and scaled distributions for all groups, and ordinal dependent variable. Effect size (proportion of variance predicted from the independent variable) was calculated as $(H-k+1)/(N-k)$, where $H$ is the Kruskal-Wallis test statistic, $k$ is the number of groups (here 3), and $N$ is the total number of observations. This effect size is called $\eta^2$, where $\eta^2=.01$ is a small effect size, $\eta^2=.06$ is a medium effect size, and $\eta^2=.14$ is a large effect size \cite{Cohen.1988}. 

\subsection{RQ3: Simulating students' preconceptions with ChatGPT}

Similar to simulating different student cohorts, it is equally possible to simulate students with different preconceptions. The FCI is a considerately designed instrument, based on an assessment of students' beliefs about force against a normative theory, namely Newtonian physics. It is recognized that overcoming preconceptions is among the crucial tasks in physics and science education, and it is a hard problem. The FCI can help instructors to align their instruction and assess students' understanding. The FCI addresses different preconceptions that students can hold. The authors identify general categories of preconception, namely: kinematics (K), impetus (I), active force (AF), action-reaction pairs (AR), concatenation of influences (CI), and other influences on motion (OI). Category K indicates if learners have a physics notion on motion description. Category I refers to learner's notion that force is some kind of intrinsic quantity to objects that keeps them moving, once it has been supplied to them. AF then refers to the conception that motion implies an active force. AR refers the learner's conception of interactions as a struggle where the stronger/more forceful party exerts greater force (most typically exemplified in the crash of a truck and a car). CI aggregates several preconceptions, such as the dominance principle in the composition of two unequal forces on a single object. Finally, also OI subsumes several preconceptions such as the reasoning with centrifugal force and considering it as a real force.

Given the well-established theory on preconceptions related to the Newtonian mechanics and forces in particular, we can utilize this information to prompt ChatGPT to generate (simulate) responses as-if it were a learner with this preconception. To do so, we deliberately engineered the prompts given to ChatGPT. We first checked if ChatGPT knew about (could elaborate on) the respective preconception (K, I, AF, AR, CI, OI). In fact, ChatGPT knew about all preconceptions and could elaborate what they were. Afterwards, we engineered a prompt where the items were preceded with the following phrase: \enquote{Respond to the following questions from the perspective of a person who has the preconception that ...}. We then described the preconception in brief or simply inserted \enquote{the impetus preconception} for agreed-upon preconception \cite{Kortemeyer.2023}. For each preconception, at least 7 different response sets (more responses were generated for preconceptions with higher observed variability in the responses) were generated. We then analyze the distribution of correct answers. Furthermore, we prompted ChatGPT for important items to provide reasoning to their responses, given that it simulated a student with a different preconception. We validated if the reasoning was in alignment with the preconception.

\section{Findings}

\subsection{RQ1: Probing conceptual understanding of ChatGPT in physics}

Regarding RQ1, we first transformed the response matrix into a binary coded matrix where correct responses were coded with 1 and incorrect responses were coded with 0. Based on this binary matrix we calculated the average correct answers (of overall $N$=14 responses) and the standard deviation. We found that $M$=0.83 ($SD$=0.22), i.e., ChatGPT scored on average 83\% in the FCI (with some transcribed items). In Table \ref{tab:statistics} the distributional statistics for all items with a mean score below 1.0 are displayed. It can be verified that for many items ChatGPT perfectly solved them. With other items (e.g., A8) there is quite some variance and solution probability is almost equal to random guessing. Item A20 was never solved correctly by ChatGPT. The engineering students ($N$=121) performed on average $M$=0.14 ($SD$=0.20).

\begin{table}[]
    \centering
\caption{Item statistics (generated with assistance of ChatGPT). Only items with a mean score below 1.00 are included.} 
\begin{tabular}{l r r r r r r}
\toprule
Item  &  Mean &  Median &   Std.dev. &  Variance &    ~~5\%~~ &   ~~95\% ~~\\ \midrule
A8  &  0.14 &    0.00 & 0.36 &      0.13 & 0.00 & 1.00 \\
A13 &  0.93 &    1.00 & 0.27 &      0.07 & 0.65 & 1.00 \\
A14 &  0.79 &    1.00 & 0.43 &      0.18 & 0.00 & 1.00 \\
A16 &  0.50 &    0.50 & 0.52 &      0.27 & 0.00 & 1.00 \\
A17 &  0.71 &    1.00 & 0.47 &      0.22 & 0.00 & 1.00 \\
A18 &  0.64 &    1.00 & 0.50 &      0.25 & 0.00 & 1.00 \\
A19 &  0.36 &    0.00 & 0.50 &      0.25 & 0.00 & 1.00 \\
A20 &  0.00 &    0.00 & 0.00 &      0.00 & 0.00 & 0.00 \\
A21 &  0.57 &    1.00 & 0.51 &      0.26 & 0.00 & 1.00 \\
A29 &  0.57 &    1.00 & 0.51 &      0.26 & 0.00 & 1.00 \\
A30 &  0.71 &    1.00 & 0.47 &      0.22 & 0.00 & 1.00 \\ \bottomrule
\end{tabular}
\label{tab:statistics}
\end{table}

\subsection{RQ2: Simulating conceptual understanding of different cohorts}

\begin{figure}
  \centering
  \includegraphics[scale=.5]{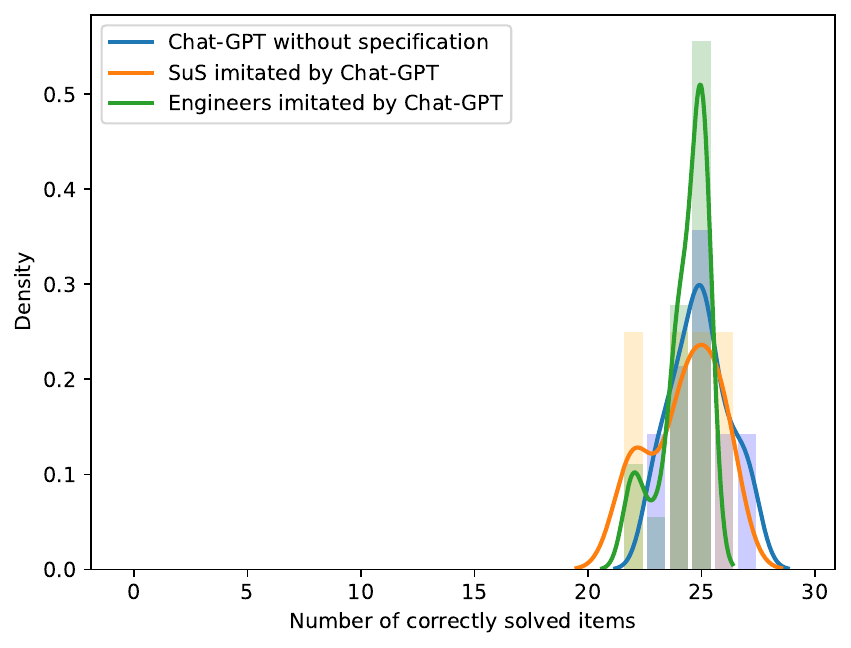}
  \caption{Distribution of correct responses for different cohorts as prompted and generated by ChatGPT.}
  \label{fig:cohort_distribution}
\end{figure}

Next, we prompted ChatGPT to answer the FCI as if it were a) a 10th grade student in school or b) an engineering student. In Table \ref{tab:desc_stats_groups}, the means and standard deviations for the distributions of the groups are depicted. In a reasonable approximation, we can verify that the scales (i.e., standard deviations) are similar for the groups, so we meet the assumptions for the Kruskal-Wallis test. This can also be visually inspected in Figure \ref{fig:cohort_distribution}, where the distributions of correct responses for the different cohorts are depicted. The Kruskal-Wallis test yields a value for potential group differences of $H$=1.79, $p$=0.409, $\eta^2$=-0.01. Essentially, no differences were found between the aggregated correct response distributions for the three cohorts.

\begin{table}[]
    \centering
\caption{Descriptive statistics for groups.\label{tab:desc_stats_groups}}
\begin{tabular}{lrrrrrr}
\toprule
{} &   Mean &  Median &   Std &  Variance &    5\% &   95\% \\
Group     &        &         &       &           &       &       \\
\midrule
Engineers &  24.28 &    25.0 &  1.02 &      1.04 &  22.0 &  25.0 \\
Unspecified     &  24.93 &    25.0 &  1.27 &      1.61 &  23.0 &  27.0 \\
Students  &  24.25 &    24.5 &  1.58 &      2.50 &  22.0 &  26.0 \\
\bottomrule
\end{tabular}
\end{table}

\subsection{RQ3: Simulating students' preconceptions with ChatGPT}

\begin{figure}
  \centering
  \includegraphics[scale=.5]{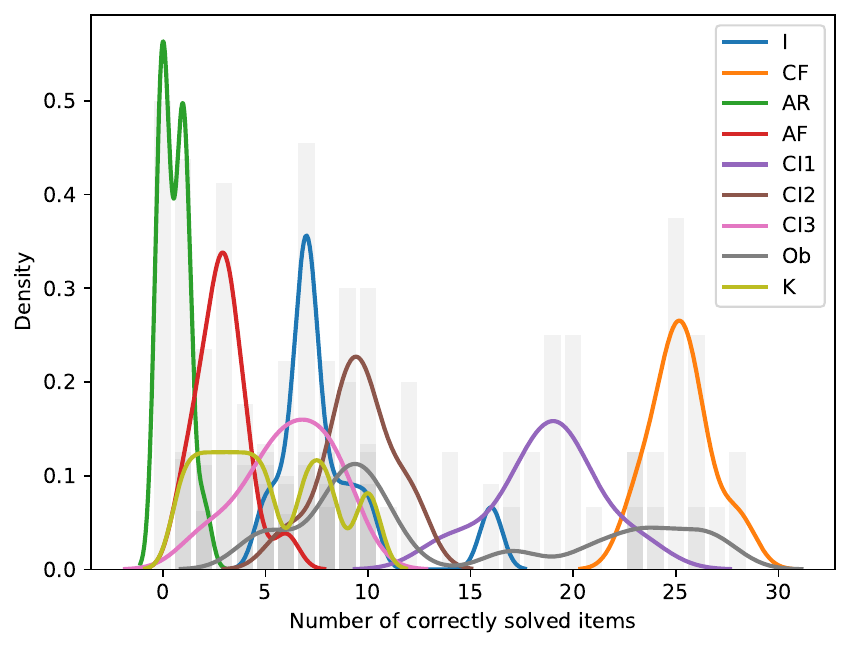}
  \caption{Distribution of correct responses for different preconceptions as prompted and generated by ChatGPT.}
  \label{fig:cohort_distribution2}
\end{figure}

\begin{figure}
  \centering
  \includegraphics[scale=.5]{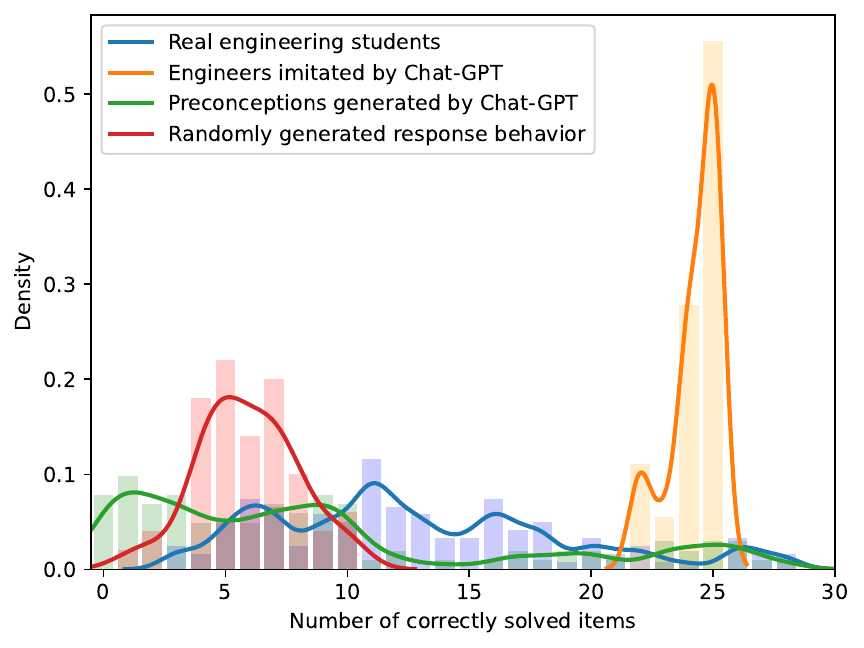}
  \caption{Distribution of correct responses for different cohorts as prompted and generated by ChatGPT.}
  \label{fig:cohort_distribution3}
\end{figure}

Figures \ref{fig:cohort_distribution2} and \ref{fig:cohort_distribution3} show the distributions for the simulated students with different preconceptions. It can be seen that they widely spread, and depending on the specific preconceptions more or less accuracy can be observed in the response sets. Interestingly, students with the AR preconception almost systematically fail at all items, whereas simulated students with CF solve the FCI as accurately as without preconceptions \ref{fig:cohort_distribution3}). Certainly, a much greater variability as in randomly generated responses can be observed.

Finally, we qualitatively validated two preconceptions with regards to the provided reasoning by ChatGPT. As for their intricate and prevalent nature, we chose centrifugal force (CF) and impetus (I). The centrifugal force is not a real force as for example gravity, but rather a pseudo force that results from the chosen frame of reference. In the inertial frame of reference, only the centripetal force acts on the object attached to the string that keeps it on the circular track. Once the string is cut, no force (except for gravity and normal force) is acting on the moving object, which, according to Newton's first law keeps on moving with constant velocity (assuming frictionless movement). The centrifugal force during a circular motion in a stationary frame of reference (also termed reactive centrifugal force, e.g., in a car) feels very real for students as they experience this force \cite{kuchemann2020students}. Centrifugal force is assumed in a center of mass frame of reference that moves alongside the object in the circular track. No net force should act on the object, as in this frame of reference it is at rest. Students with this preconception would choose answer 4, i.e., after the string is plugged, the outward force (centrifugal force) will drive the object outwards. ChatGPT reasons this case convincingly and correctly. It is noteworthy that no mention of the frame of reference problem is made.

\begin{displayquote}
\textbf{Prompt}, (see Section \ref{sec:methods}):

\textbf{Responses by ChatGPT (translated by ChatGPT):} First response:
From the perspective of a person who incorrectly believes that a centrifugal force exists, they may choose Answer 4: \enquote{A trajectory that initially goes from the center outwards, but then takes a trajectory in the direction of the original movement.}

The misconception of centrifugal force is based on the feeling that a force is acting during circular motion that pushes the body outwards. In reality, it is the centripetal force (a force directed inwards) that keeps the body on its circular path. This misconception could lead a person to believe that when the string breaks, the ball is initially driven outwards from the center (by the supposed centrifugal force) before then moving in the direction of the original motion because there is no longer any centrifugal force to push it outwards.

The correct answer would, however, be Answer 2: \enquote{A linear trajectory that shows tangential along the circle at the breaking point.} Because when the string breaks, there is no longer a centripetal force that forces the ball into a circular path, and the ball continues its motion in a straight line, which is tangential to the original circular path, due to inertia. This motion corresponds to Newton's first law, which states that a body remains at rest or moves with constant velocity in a straight line unless a force acts on it.
\end{displayquote}

As for the impetus concept, students typically believe (much in agreement with their lived experiences in a friction-based world) that when an object is thrown, something is transferred to the object that it can use to fly a certain time, constantly using up the impetus. Once it is used up, the objects comes to rest. We would expect students to choose answer 1 for this item. After all, a circular impetus is given to the object (e.g., by the string). Once the string is plugged, this circular impetus will cause the object to continue on the circular path. Again, ChatGPT correctly reasoned about this preconception and chose the correct answer, responding as-if it were a student with the impetus preconception.

\begin{displayquote}
\textbf{Prompt}, (see Section \ref{sec:methods}):

\textbf{Responses by ChatGPT (translated by ChatGPT):}\
Second response:
The theory of impetus is a medieval physical theory that was proposed before Newton's laws of motion. This theory suggests that a moving object possesses an \enquote{impetus} that drives it further in the direction of the original motion, even when no external force acts on it anymore. It is believed that the impetus gradually decreases, and the object eventually stops.

With this misconception, a person might choose Answer 1: \enquote{A trajectory that closely follows the course of the circular path.} This person might believe that after the string breaks, the ball continues to fly in the original circular motion because the impetus drives it in that direction. The ball would then gradually deviate from the circular path and finally stop when the impetus has completely diminished.

However, the correct answer according to modern laws of physics would be Answer 2: \enquote{A straight-line trajectory that is tangential to the circle at the breaking point.} According to Newton's first law of motion, in the absence of an external force (which was removed by the breaking string), the ball would continue its motion in a straight line, which is tangential to the circular path it was originally on.
\end{displayquote}

\section{Discussion}

The recent advances in AI research, in particular generative AI, provide novel potentials for physics education researchers to enhance research and teaching, however, they also introduce novel challenges. To better understand the potentials and challenges, it is crucial to engage with the advanced technologies, such as LLM, to evaluate their performance for physics education-specific problems (as argued by \cite{Wang.2023}, who called for \enquote{more careful and controlled studies} on this matter, particularly with ChatGPT). In this study, we evaluated the capabilities of an LLM, namely ChatGPT, to generate synthetic data for the FCI and thus potentially enhance concept inventory development. We chose the FCI as a use case as this is a well validated and most widely used concept inventory in physics. Prior studies established that ChatGPT is well versed to solve the FCI. In our study, we advanced this research to compare performance with real students (RQ1), and generate synthetic data for it with regards to different cohorts (RQ2), and varying preconceptions (RQ3).

In RQ1 we found that ChatGPT (based on GPT 4) solved the FCI (with transcribed images) accurately. In fact, it solved the FCI much more accurately compared to real engineering students in a German university. This resonates with prior studies that established that ChatGPT can accurately solve the FCI \cite{Kortemeyer.2023,west2023ai}. Compared to the prior work by West, where ChatGPT (based on GPT-4) solved 22 out of 23 items correctly, here ChatGPT (based on GPT-4) only correctly solves about 26 out of 30 items on average. There were eleven items that have not been solved correctly in each run (see Tab. \ref{tab:statistics}). Six of these eleven items contained a figure that was translated into text. In comparison, four items that contained text translated from a figure were solved correctly in each run. Consequently, the translation from figures to text may cause difficulties for ChatGPT to interpret. Previously, some of these items were also taken out from consideration \cite{west2023ai}. As a note of interest, ChatGPT can be easily prompted to output the responses in varying formats (e.g., comma-separated list), and it solves it much more quickly than real students. We particularly generated more response sets in order to also retrieve information on the variance displayed in responding to the FCI. For most items, no variance was found, because they were solved with 100\% accuracy. For other items, however, a mere random performance was observed.

In RQ2 we sought to utilize ChatGPT (via specific prompting) to simulate answers from students in different cohorts (10th grade students, and engineering students). To our surprise, we found no differences between these cohorts and the unprompted (with regards to cohort) responses. Either the mean accuracy was similar, and the standard deviation was similar. This struck us as unexpected, given that it is established that prompting can in fact be used to modify the responses of ChatGPT and other generative AI-based LLM \cite{Liu.2021}. However, maybe the provided prompts were not specific enough to affect the responses in ChatGPT. 

Therefore, in RQ3, we specifically prompted ChatGPT to respond to the FCI as-if it were a student with a certain preconception related to Newtonian force concept and mechanics more generally. Here, we found large differences in the response behaviour of ChatGPT. ChatGPT knew about all preconceptions. The accuracy, given that ChatGPT responses as-if it were a student with almost any preconception, dropped considerably, sometimes even below random guessing performance. Interestingly, prompting ChatGPT with preconceptions resulted in distributions that were equally spread as real students’ distributions. For two preconceptions we verified that ChatGPT also correctly reasoned if it was prompted to respond to the items as-if it had a certain preconception, which it did. 

These data augmentation techniques could be of interest to physics education researchers and educational researchers more generally if they want to pilot test their instrument, e.g., in the domain of mechanics. Even practitioners designing classroom exams and policy agents designing state-wide exams could benefit from these capabilities of generative AI (here: LLM) to assure validity of their instruments. One caveat would be that users currently will have to outsource their instruments to a private company. On the other hand, it can be tiresome to students to function as “lab animals” and be subjected to pilot testing of research instruments (let alone ethical concerns). Moreover, precious time of the students is consumed which could be used in more learning-relevant activities.

\section{Limitations}

While LLM such as GPT 4 provide novel capabilities for research such as data augmentation, they also introduce novel challenges that researchers have to engage with. Brown et al. \cite{Brown.2020} list three: misuse, bias and energy consumption as important points of concern. For once, LLM such as GPT 4 can be used to generate human-sounding text and therefore be used to spread certain narratives. With regards to physics education research, students might fabricate their essays with ChatGPT, which cannot be easily detected by human instructors or researchers \cite{Yeadon.2023}. In the future, the selection of teaching materials should be carefully tailored to ensure that tools like ChatGPT are not undesirably misused as aids. For example, we can observe a similar context where students increasingly resort to aids such as YouTube videos when the provided materials exhibit deficiencies \cite{PhysRevPhysEducRes.16.020123}. Equally concerning is the problem of hallucination in LLM such as GPT 4. For example, ChatGPT provides well-sounding output with sometimes flawed physics \cite{Gregorcic.2023}. This becomes a problem when considering that students accept the outputs of Chat-GPT without critical reflection (copy and paste it) \cite{unknown}. Instructors need to find means to prevent tasks where simple copy paste is sufficient. We also found that important steps in the problem solving process such as explicating on underlying assumptions and idealizations are oftentimes omitted \cite{Kieser.inpress}. This makes it difficult for students to truly learn the underlying physics. Kortemeyer \cite{Kortemeyer.2023} prompted ChatGPT on its ability to understand physics and received the following response:

\begin{displayquote}
\enquote{as a language model, I have been trained on a large dataset of text, including physics texts. This allows me to understand and generate text related to physics concepts, but it does not mean that I have the ability to solve physics problems or pass a physics course. I can provide explanations and answer questions about physics to the best of my knowledge, but I am not a substitute for a human physics expert or a physics education.}
\end{displayquote}

As such, human expert judgements will remain crucially important to verify responses given by ChatGPT. As a matter of fact, we found the theory-based, specific prompting with regards to established preconceptions yielded most variance which could be important for data augmentation. As such, established theory is crucial to guide prompt engineering and generate meaningful synthetic data.
It was furthermore established that LLM such as GPT 3 and others incorporate human-like biases, e.g., with regards to gender \cite{Caliskan.2017,Brown.2020}. Given the fact that they are trained on the Common Crawl of the Internet and similar unclean data, this is probably not surprising. However, approaches to mitigate this problem are still not fully developed and thus employing ChatGPT and other LLM in educational settings should probably be postponed, especially in contexts of high-stakes testing. Finally, energy consumption of training these LLM and even requesting a single output are ecologically challenging \cite{Dodge.2022}. It will likely be not sustainable for every research group to run their own models. Shared infrastructures will be required to make these applications sustainable.

\section{Implications}

Our study showed that with adequate prompting strategies ChatGPT can be utilized to generate synthetic data related to a concept inventory with meaningful variability in the data set. We submit that similar conclusions will likely be true for other concept inventories as well, such as climate change, electricity, and also biological evolution. Further research is needed to test these predictions. This offers researchers a valuable tool to simulate field testing their concept inventories and evaluate psychometric properties, given sufficient sample sizes. We could show that similar distributions as with real students emerge, and further research could investigate if convolutions of preconceptions will enable a reconstruction of the true students’ distribution. We cannot, however, exclude the possibility that real students also show other response patterns, i.e., have different conceptions that are not included in our theoretically derived set of preconceptions. Further research should investigate also the possibility to even generate a concept inventory in the first place with the help of LLM such as GPT 4 \cite{Kuchemann.2023}, and assist in other phases and steps of the development process for a concept inventory \cite{Adams.2011}.

\subsection*{Data and code availability:}

The code for analysis (and further analyses) can be accessed here:
\url{https://colab.research.google.com/drive/1WU_cBfcYMO2dxE9kFVjK9Hlq0mSati8V?usp=sharing}

The synthetic data can be accessed upon reasonable request to the authors.

\subsection{Translation of figures in FCI items} \label{}
In this section, we show the translation of figures in FCI items into text so that ChatGPT was able to interpret it. We did not include the original figures due to copyright restrictions, but the item numbering is identical to the original work \cite{hestenes1992force}.

{\bf Item 5}

Which of the following forces act(s) on the ball when it is exactly halfway through a semicircular channel?
\begin{enumerate}
\item[A)] A downward force of gravity
\item[B)] A force exerted by the channel that is directed from the point where the ball is currently located to the center of the channel
\item[C)] A force in the direction of motion
\item[D)] A force directed from the center point to the point in the channel where the ball is currently located.
\end{enumerate}
\begin{enumerate}
    \item force A only
    \item forces A and B
    \item forces A and C
    \item forces A, B, and C
    \item forces A, C, and D
\end{enumerate}

{\bf Item 6}

Which path would a ball most closely follow after it exits the channel at the end and moves across the frictionless table top?
\begin{enumerate}
    \item A path that would correspond to the arc of the gutter if it formed a closed circle.
\item A straight line trajectory curve that points tangentially along the circle at the exit point.
\item A trajectory curve that describes an arc of a circle outwards
\item A path curve which describes a parabola to the outside
\item A trajectory in which the ball is immediately deflected outward in a straight line

\end{enumerate}

{\bf Item 7}

A steel ball is attached to a string and is swung in a circular path in a horizontal plane. At one point, the string suddenly breaks near the ball. If these events are observed from directly above, which path would the ball most closely follow after the string breaks?
\begin{enumerate}
    \item A trajectory that almost corresponds to the course of the circular path.
    \item A straight-line trajectory that points tangentially along the circle at the tear-off point.
    \item A straight-line trajectory that points away from the circular path at an angle of 45° from the tangent line at the break-off point.
    \item A trajectory that initially goes outward from the center, but then takes a trajectory in the direction of the original motion.
    \item A trajectory that points outward in a straight line from the center.

\end{enumerate}

{\bf Item 8}

Imagine a hockey puck is sliding with constant speed $v_0$ in a straight line from a point "P" to point "Q" on a frictionless horizontal surface. Forces exerted by the air are negligible. You are looking down on the puck. When the puck reaches point "Q," it receives a swift horizontal kick perpendicular to the previous direction of motion. Had the puck been at rest at point "Q," then the kick would have set the puck in horizontal motion with a speed $v_k$ in the direction of the kick.

Which paths would the puck most closely follow after receiving the kick?
\begin{enumerate}
    \item A straight line in the direction of the collision
\item A straight line pointing at a sharp angle to the previous trajectory 
\item A path in the same direction as the original direction of motion but offset in parallel in the direction of the impact
\item A curved path that slopes in the direction of the shock
\item A curved path inclined in the direction of the original direction of motion
\end{enumerate}

{\bf Item 12}

A ball is fired by a cannon from the top of a cliff in horizontal direction. Which paths would the cannon ball most closely follow?
\begin{enumerate}
    \item A path that runs in a straight line down to the point of impact.
    \item A path that is curved toward the point of impact.
    \item A path which is curved towards the point of impact, but which is horizontal for a short distance after launch.
    \item A path that is curved towards the point of impact, but is horizontal for a certain distance after launch and vertical  before impact.
    \item A path that runs horizontally until almost vertically above the point of impact and then runs vertically until the point of impact.

\end{enumerate}
{\bf Item 14}

A bowling ball accidentally falls out of the cargo bay of an airliner as it flies along in a horizontal direction. As observed by a person standing on the ground and viewing the plane from the side, which path would the bowling ball most closely follow after leaving the airplane?
\begin{enumerate}
    \item A curved path that runs backwards.
\item A path that runs vertically downward.
\item A path that runs forward in a straight line.
\item A curved path that runs forward.
\item A path that extends forward horizontally to above the point of impact and then vertically downward.
\end{enumerate}

{\bf Items 15, 16 and 17}

The figures have not been described as they only carry redundant information that is also contained in the text. 

{\bf Item 18}
A child swings on a rope tied to a branch (point O). His movement starts above a point P and then continues downward through the point P. 
Consider the following distinct forces:
\begin{enumerate}
    \item A downward force of gravity.
    \item A force exerted by the rope pointing from A to O.
    \item A force in the direction of the boy’s motion.
    \item A force pointing from O to A.
\end{enumerate}
Which of the above forces is (are) acting on the boy when he is at position A?
\begin{enumerate}
    \item[(A)] 1 only.
    \item[(B)] 1 and 2.
    \item[(C)] 1 and 3.
    \item[(D)] 1, 2, and 3.
    \item[(E)] 1, 3, and 4.
\end{enumerate}

{\bf Item 19}

Two blocks move to the right. The following table shows the distance covered by the blocks at intervals of 0.2 s in length units (LU). Identical numbers indicate identical points in time.

Block a: 1: 2 LU, 2: 4 LU, 3: 7 LU, 4: 11 LU, 5: 16 LU, 6: 22 LU, 7: 29 LU

Block b: 1: 0 LU, 2: 4 LU, 3: 8 LU, 4: 12 LU, 5: 16 LU, 6: 20 LU, 7:24 LU, 8: 28 LU

Do the blocks ever have the same speed?
\begin{enumerate}
    \item No.
    \item Yes, at instant 2.
    \item Yes, at instant 5.
    \item Yes, at instants 2 and 5.
    \item Yes, at some time during the interval 3 to 4.
\end{enumerate}

{\bf Item 20}

Two blocks move to the right. The following table shows the distance covered by the blocks at intervals of 0.2 s in length units (LU). Identical numbers indicate identical points in time.

Block a: 1: 2 LU, 2: 6 LU, 3: 10 LU, 4: 14 LU, 5: 18 LU, 6: 22 LU, 7 26 LU

Block b: 1: 0 LU, 2: 6 LU, 3: 12 LU, 4: 18 LU, 5: 24 LU

The accelerations of the blocks are related as follows:
\begin{enumerate}
    \item[(A)] The acceleration of ``a" is greater than the acceleration of ``b".
    \item[(B)] The acceleration of ``a" equals the acceleration of ``b". Both accelerations are greater than zero.
    \item[(C)] The acceleration of ``b" is greater than the acceleration of ``a".
    \item[(D)] The acceleration of ``a" equals the acceleration of ``b". Both accelerations are zero.
    \item[(E)] Not enough information is given to answer the question.
\end{enumerate}

{\bf Item 21}

A rocket drifts sideways in outer space from point "a" to point ``b" as shown below. The rocket is subject to no outside forces. Starting at position ``b", the rocket's engine is turned on and produces a constant thrust (force on the rocket) at right angles to the line ``ab". The constant thrust is maintained until the rocket reaches a point ``c" in space.

Which path below best represents the path of the rocket between points ``b" and "c"?
\begin{enumerate}
    \item Point ``c" lies to the right above point ``b". The path from ``b" to "c" is concave. 
    \item Point ``c" is right above point ``b". The path from ``b" to ``c" is straight and perpendicular to the line connecting ``b" to ``a". 
    \item Point ``c" lies to the right above point ``b". The path from ``b" to ``c" is straight.  	
    \item Point ``c" lies above and to the right of point ``b". The path curve from ``b" to ``c" runs horizontally to the right at the beginning and is convex from then on.
    \item Point ``c" lies above and to the right of point ``b". The path curve from ``b" to ``c" is convex.
\end{enumerate}

{\bf Item 23}

At point ``c" the rocket's engine is turned off and the thrust immediately drops to zero.

Which of the paths below will the rocket follow beyond point ``c"?
\begin{enumerate}
    \item The path curve runs in a straight line horizontally to the right. 
    \item The path curve runs in a straight line diagonally upwards to the right. 	
    \item The path curve runs vertically upwards in a straight line. 	
    \item The path curve runs to the right and is concave. 	
    \item The path curve runs to the right and is convex.
\end{enumerate}

{\bf Item 28}
The figure has not been described as it only carries redundant information that is also contained in the text. 
    

\bibliography{cas-refs}



\end{document}